\documentclass[preprint,12pt]{elsarticle} 

\usepackage{amssymb}
\usepackage{amsthm}
\usepackage{amsmath,amssymb,amsfonts}%
\usepackage{multirow}%
\usepackage{tabularx}
\usepackage{caption}
\usepackage{float}
\usepackage{multicol}
\usepackage{multirow}
\usepackage{booktabs}
\usepackage{caption}
\usepackage[utf8]{inputenc}
\usepackage[T1]{fontenc}
\usepackage{xcolor}
\usepackage{soul}
\usepackage{listings}%
\usepackage{threeparttable}
\usepackage{textcomp}%
\usepackage{xcolor}%
\usepackage{mathrsfs}%
\usepackage[polish,english]{babel} 
\usepackage[colorlinks,citecolor=red,urlcolor=blue,bookmarks=false,hypertexnames=true]{hyperref}

\journal{Computational Materials Science}

\begin{document}
\emergencystretch 3em
\begin{frontmatter}

\title{Machine-learned accelerated discovery of oxidation-resistant NiCoCrAl high-entropy alloys}

\author[a]{Dennis Boakye}
\author[a]{Chuang Deng \corref{b}}
\cortext[b]{Corresponding author}
\ead{chuang.deng@umanitoba.ca}

\affiliation[a]{organization={Mechanical Engineering, University of Manitoba},
            addressline={66 Chancellors Cir}, 
            city={Winnipeg},
            postcode={R3T 2N2}, 
            state={Manitoba},
            country={Canada}
            }

\begin{abstract}
The development of oxidation-resistant high-entropy alloy (HEA) bond coats is restricted by the limited understanding of how multi-principal element interactions govern scale formation across temperatures. This study uncovers new oxidation trends in NiCoCrAl HEAs using a data-driven analysis of high-fidelity experimental oxidation data. The results reveal a clear temperature-dependent transition between alumina- and chromia-dominated protection, identifying the compositional regimes where alloys rich in Al dominate at $\ge1150$ °C, mixed Al–Cr chemistries are optimal at intermediate temperatures, and, unexpectedly, Cr-rich low-Al alloys perform best at 850 °C—challenging the assumption that high Al is universally required. The effects of Hf and Y are shown to be strongly composition-dependent with Hf producing the largest global reduction in oxidation rate, while Y becomes effective primarily in NiCo-lean alloys. Y–Hf co-doping offers consistent improvement but exhibits site-saturation behavior. These insights identify new high-performing HEA bond-coat families, including $\mathrm{Ni_{17}Co_{23}Cr_{30}Al_{30}}$ as a substitute for conventional mutlilayer thermal barrier coatings.
\end{abstract}




\end{frontmatter}


\section{Introduction} \label{intro}
Advanced structural materials, especially those based on Ni or Co superalloys, are fabricated to withstand extreme mechanical loads and resist creep at high temperatures \cite{selvaraj2021recent,bauer2012creep}. Despite continuous advances, their performance at elevated temperatures is flawed by the phase stability and diffusion kinetics of their constituent elements \cite{kazantseva2018superalloys}. Moreover, as pressures and operating temperatures continue to increase in pursuit of higher thermal efficiency, superalloys are pushed closer to their limits, leading to phase instability \cite{tin2003phase}, grain boundary degradation \cite{suzuki2019degradation}, hot corrosion \cite{patel2017high}, and accelerated oxidation \cite{pint2010oxidation}. These collective effects compromise structural integrity and shorten the life of components, leading to catastrophic failures. Furthermore, the composition of many superalloys includes critical raw materials such as rhenium and tantalum, whose availability and cost pose an additional constraint on their large-scale usage \cite{yeh2008creep,pollock2006nickel,sun2014effect,zhu2025cyclic}.

To solve these issues, the application of special coating materials has become an integral part of protecting structural components from extreme thermal and oxidative environments \cite{guo2023thermal,stiger1999thermal,padture2002thermal}. A typical thermal barrier coating (TBC) is yttria-stabilized zirconia (YSZ), which is composed of a ceramic top coat and a metallic bond coat over a superalloy substrate. The top coat helps mitigate exposure to temperature, while the bond coat provides resistance to oxidation against oxygen atoms that diffuse through the top coat \cite{vv2019role,borom1996role}. For the past 30 years, YSZ has remained the industry standard due to its low thermal conductivity, good mechanical strength, and phase stability up to 1200$\mathrm{\,^\circ C}$ \cite{padture2002thermal,diwahar2025overview}. However, it undergoes a transition from tetragonal-to-monoclinic phase at higher temperatures that causes microcracks and spallation during thermal cycling \cite{liu2016effect,vassen2013thermal}. Moreover, the introduction of vertically cracked and columnar microstructures \cite{mehboob2021tailoring} to counter these issues remains insufficient, especially under severe cyclic conditions \cite{zhu2000thermal,ganvir2018tailoring,pakseresht2022failure}. Another key failure mechanism in TBC systems is the growth of a thermally grown oxide (TGO), usually $\mathrm{\alpha}$-$\mathrm{Al_2O_3}$, at the interface between the bond coat and the top coat, which thickens over time and generates stresses that lead to interfacial delamination \cite{clarke2003materials,evans2011oxidation}. This degradation worsens when trace impurities, such as sulfur, migrate from the bulk to the metal-oxide interface and weaken interfacial bonds \cite{bai2016migration,bose2017high,evans2011oxidation,sarioglu1996control,liu2023comparative}. As a result, new materials with superior thermal and oxidation resistance are being sought.

In the wake of multi-principal element alloying, high-entropy alloys (HEAs) show promising alternatives to TBCs in high-temperature applications \cite{sharma2021high,qiu2014effect,cheng2014effect}. The high configurational entropy promotes the formation of simple solid-solution phases with exceptional thermal stability, oxidation resistance, and mechanical strength \cite{tsai2014high,garg2023improving,vaidya2019phase}. HEA-fabricated coatings can serve as bond coats and top coats, potentially simplifying the multilayer structure of conventional TBC systems \cite{shahbazi2023high}. Miracle et al. \cite{miracle2014exploration} and Oleksak et al. \cite{oleksak2024high} demonstrated that selected HEAs formed slow-growing adherent oxides that outperformed conventional Ni-based alloys in high-temperature oxidation. Moreover, the compositional tunability of HEAs allows for tailoring properties such as thermal expansion \cite{huang2017thermal,zhang2024effect,lin2021investigation}, oxidation behavior \cite{holcomb2015oxidation,lu2020hf}, and resistance to volatile species \cite{zhang2022assessment,dehury2024elemental}, making them ideal for environments demanding thermal protection and resistance to chemical degradation. However, the overarching problem is the expansive design space of HEAs \cite{cantor2004microstructural}. At 2$\mathrm{\,at\%}$ intervals, a single quinary HEA system contains more than 300,000 unique compositions, making traditional experimental methods or conventional computational brute-force searches impractical. Computational tools such as density functional theory and CALPHAD remain a bottleneck despite predictions of phase stability and oxidation thermodynamics because of their computational cost. \cite{feng2018phase,pandey2022theoretical,chen2022phase,ma2015phase,arshad2024high,shaburova2021high}.

Recently, the incorporation of machine learning (ML) frameworks in materials science has gained strength, particularly in predicting oxidation resistance within high-performance materials \cite{duan2023design,gorsse2025advancing,wei2023discovering}. The use of data-driven methods to establish correlations between experimental parameters and oxidation kinetics is well documented in the literature. For example, Li et al. \cite{li2024machine} used neural networks and gradient-boosting regression models to predict the oxidation kinetics in high temperature oxidation of HEAs based on temperature, time and elemental compositions. Dong et al. \cite{dong2023machine} combined an experimental study and a random forest regression model to classify the compositions of HEAs resistant to oxidation. The researchers obtained a high precision between the experimental findings and the model-predicted values. Khatavkar and Singh \cite{khatavkar2024combined} used artificial neural networks and tree-based models to study the oxidation kinetics of Ni superalloys. The models predicted alloys with high oxidation resistance, emphasizing the role of machine learning in high-throughput screening. Recently, Tan et al. \cite{tan2025machine} applied gradient-boosting regression to predict oxidation resistance in nonequiatomic NiCoCrAlFe HEAs, combining machine learning with CALPHAD thermodynamic calculations to propose alternatives for oxidation-resistant bond coats. 

NiCoCrAl alloys are widely used as bond coats due to their superior oxidation resistance and thermal stability at elevated temperatures \cite{chen2015characterization,hemker2008characterizing,seraffon2014oxidation}. Moreover, they facilitate strong adhesion between the substrate and the topcoat, enhancing the durability of TBC systems. However, existing studies on NiCoCrAl alloys focus mainly on their superalloys \cite{liu2022design,cojocaru2022nicocralx,ebach2021lifetime}. Seraffon et al. \cite{seraffon2014oxidation} demonstrated that optimal NiCoCrAl compositions can form protective scales of $\mathrm{Al_2O_3}$ or $\mathrm{Cr_2O_3}$ at elevated temperatures, significantly improving the performance of the bond coat under industrial gas turbine conditions. However, unique chemistry and oxidation pathways are often neglected due to the high sensitivity of the substrate to the balance of oxide-forming elements such as Al and Cr \cite{chen2025effects}. Although Al promotes the formation of a protective $\mathrm{\alpha}$-$\mathrm{Al_2O_3}$ scale, Cr can accelerate the formation of transient oxides and affect long-term scale adherence \cite{stott1995influence,heinonen2011initial}. Understanding the impact of Al content on phase stability is crucial for designing oxidation-resistant NiCoCrAl HEA bond coats. Liu et al. \cite{liu2022design} combined machine learning and Calphad to study the eutectic composition of NiCoCrAl HEA by varying the Al concentration. The researchers observed that alloys poor in Al consisted of an FCC ($L1_2$) phase, while those rich in Al had a BCC (B2) phase. Despite this, the vast nonequiatomic compositional space of NiCoCrAl alloys is largely uncharted in current ML models. In addition, the beneficial effects of reactive elements (REs) such as Y and Hf, which are known to improve scale adhesion and suppress spallation \cite{cojocaru2022nicocralx,lu2020hf,huang2025oxidation}, are rarely included in training datasets. Existing models also overlook the cyclic oxidation and interfacial degradation that dominate bond coat failure in service environments \cite{evans2011oxidation}, and often rely on empirical descriptors without incorporating physically meaningful features related to oxidation mechanisms.

To address these challenges, we introduce a machine learning framework tailored to predict the oxidation resistance of candidate NiCoCrAl HEA coatings across the entire compositional range. The model is trained on experimental datasets with compositions that incorporate RE additions to reflect realistic bond-coat chemistries \cite{huang2025oxidation}. The model targets the parabolic rate constant ($k_p$), which is a measure of resistance to oxidation \cite{wagner1952theoretical}. In contrast to previous models, our model captures the role of both base and REs in scale formation and degradation, enabling accurate predictions not only for alloy screening but also for understanding composition-driven mechanisms in oxidation behavior. 

\section{Material and methods}
\subsection{Data collection and processing}
\begin{figure}[h]
	\centering
	\includegraphics[width=1\linewidth]{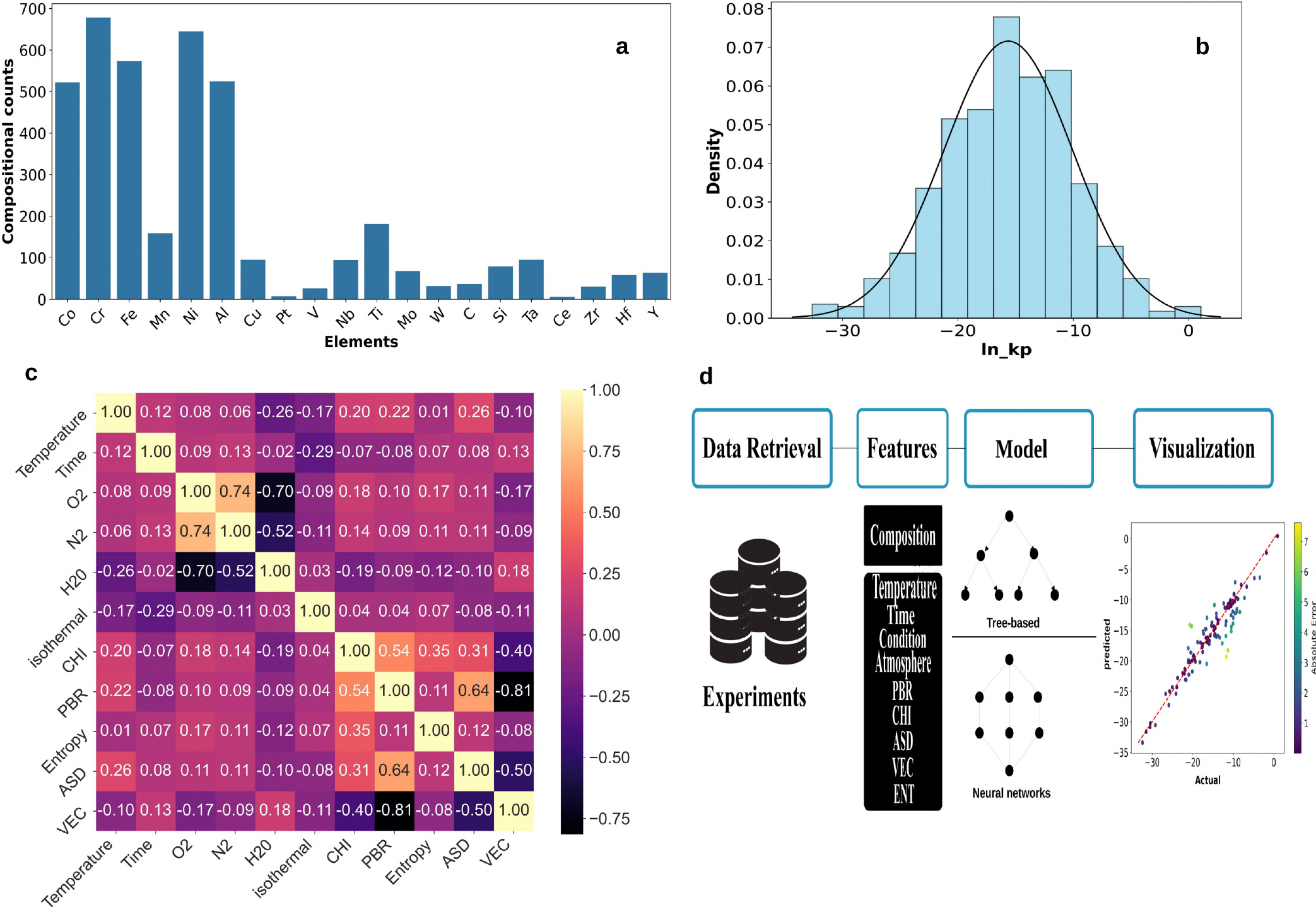}
	\caption{(a) Compositional counts of each element in the final dataset; (b) Gaussian distribution of $\ln k_p$ across the dataset; (c) Pearson coefficient matrix calculated for descriptors. Note: alloy components are excluded; (d) Model architecture development.}
	\label{fig:samp_mod}
\end{figure}

The dataset used in this study was meticulously extracted from 743 experimental measurements of oxidation behavior in HEAs, with a minor fraction originating from superalloys. The compositional diversity of the dataset is illustrated in Fig. \ref{fig:samp_mod}(a), which shows the frequency of occurrence for each element in all alloy compositions. The dataset comprising 20 elements is dominated by transition metals such as Co, Cr, Fe, and Ni, which are frequently used in oxidation studies, but also includes elements such as Al, Ti, and refractory additions such as Hf and Y, which are known to influence oxide formation and stability \cite{lu2020hf,dkabrowa2021oxidation,holcomb2015oxidation}. This compositional space ensures that the ML model is exposed to a wide range of oxidation behaviors, from fast-growing nonprotective oxides to slow-growing adherent scales. The oxidation atmospheres were represented by the respective partial pressures of the gases ($\mathrm{N_2}$, $\mathrm{O_2}$ and $\mathrm{H_2O}$) depending on the description of the oxidation environment if not explicitly stated. 

According to Wagner theory of oxidation, the kinetic curve of alloys resistant to oxidation generally follows the parabolic law, which is expressed as the ratio of the square of the weight gain per unit area to the exposure time given by \cite{wagner1952theoretical}:
\begin{equation}
	\left(\frac{\Delta m}{S}\right)^2 = k_pt + C\label{eqn1}
\end{equation}
where $\Delta m$ is the weight gain, $S$ is the surface area of the alloy, $t$ is the exposure time and $C$ is a fitting constant. Data were taken exclusively from studies where $k_p$ was reported. In cases where only the mass gain was provided, $k_p$ was obtained by dividing the square of the mass gain by the oxidation time, ensuring a consistent kinetic measure across all entries. To compress the dynamic range given that $k_p$ and exposure time span several orders of magnitude between different alloys and testing conditions, a natural logarithmic transformation ($\ln k_p$) was applied to improve numerical stability during model training. Fig. \ref{fig:samp_mod}(b) shows the resulting $\ln k_p$ distribution, which exhibits a normal distribution, indicating the suitability for regression-based machine learning methods. Units of exposure time and temperature were converted to hours and $^\circ C$, respectively. The target variable, $\ln k_p$, was converted to a standard unit of $\mathrm{mg^2cm^{-4}s^{-1}}$.

The experimental parameters associated with each oxidation measurement were complemented by incorporating a set of oxidation-linked physical principles to capture intrinsic material tendencies relevant to oxidation kinetics. These include: (i) the Pilling–Bedworth ratio (PBR) which quantifies the volumetric relationship between oxide and substrate to assess scale protectiveness; (ii) atomic size difference (ASD), which influences lattice distortion and diffusion pathways; (iii) the electronegativity difference ($\chi$), reflecting chemical reactivity and the driving force for selective oxidation; (iv) configurational entropy ($\Delta S_{conf}$), linked to phase stability and suppression of detrimental multiphase oxidation; and (v) the valence electron concentration (VEC), which affects phase formation and bonding. These descriptors, although previously applied in corrosion studies \cite{zeng2024machine,sasidhar2023enhancing}, are equally relevant to oxidation because both phenomena involve coupled transport and interfacial chemical reactions \cite{tan2025machine}. The definitions for these descriptors are provided in Table S1 of the Supplemental File.

To retain distinct physical contributions and ensure that highly collinear features do not bias the model, Pearson's correlation analysis was performed to identify potential redundancies. A correlation coefficient of 1 indicates a perfect positive correlation, whereas -1 indicates a perfect negative correlation \cite{li2024machine}. The correlation map as shown in Fig. \ref{fig:samp_mod}(c) reveals, for example, a moderate coupling between the partial pressure of oxygen and the nitrogen content and a strong negative correlation between PBR and VEC, reflecting the underlying periodic trends. This analysis informed the final selection of features, ensuring that the set was comprehensive and minimally correlated, improving the robustness and predictive accuracy of the model. The final data for each entry include the atomic fraction of the alloy components, temperature, exposure time, partial pressure of the gases ($\mathrm{N_2}$, $\mathrm{O_2}$, $\mathrm{H_2O}$), oxidation condition (isothermal or cyclic), PBR, ASD, $\chi$, VEC and $\Delta S_{\text{conf}}$. Cross-correlation between elemental composition, experimental parameters and physical principles is provided in Fig. S4 of the Supplemental file.

\subsection{Model Architecture and training}
To explore the relationship between different alloy combinations, experimental parameters, and oxidation kinetics, we employed five machine learning algorithms comprising a deep neural network (DNN), random forest regression (RF), gradient boosting regression (GBoost), AdaBoost regression, and extreme gradient boosting regression (XGBoost) as shown in Fig. \ref{fig:samp_mod}(d). Each model was trained using the same dual input feature representation with normalized elemental compositions. Feature sets excluding $\ln k_p$ were independently scaled using standardization pipelines to ensure numerical stability and comparability between models. 

The DNN was implemented as a two-branch architecture in which the elemental composition and descriptor inputs were concatenated and passed through a sequence of dense layers (256, 128, and 24 neurons) with rectified linear unit and exponential linear unit activation functions. A ridge regression weight regularization and a 20\% dropout rate were applied between layers to mitigate overfitting. The output layer consisted of a single linear neuron to predict $\ln k_p$. The network was trained with Adam Optimizer with a learning rate of 0.001. Early stopping and learning rate scheduling was employed to improve the convergence stability of the model.

The tree-based ensemble models were configured with 200 tree estimators and tuned to maximum depths to balance the complexity and generalization of the model. These algorithms were selected and compared to DNN by leveraging their ability to capture non-linear feature interactions and rank variables' importance without requiring extensive feature transformations. All models were trained using the same data split, where 85\% of the dataset was used for training and the remaining 15\% for testing. This consistent split allowed for a fair comparison between the different algorithms and ensured that the predictive framework was not based on a single type of model. 

The performance of the five selected models was assessed using standard statistical measures that quantify both the accuracy and consistency of the predictions compared to the experimental values. The coefficient of determination ($\mathrm{R^2}$) and the root mean squared error (RMSE the same unit as $\ln k_p$) were used to benchmark the models. The $\mathrm{R^2}$ metric that determines the proportion of variance in the experimental values explained by the model is given by the following equation.
\begin{equation}
	R^2 = 1 - \frac{\sum_{i=1}^{n}\left(y_i - \hat{y_i}\right)^2}{\sum_{i=1}^{n}\left(y_i-\bar{y}\right)^2}
\end{equation}
where $y_i$ denotes the experimentally measured value of $\ln k_p$ for the $i$th sample, $\hat{y_i}$ is the corresponding predicted value from the model, and $\bar{y}$ is the mean of the experimental values in the dataset. The total number of data points is denoted by $n$.

The RMSE quantifies the square root of the average squared difference between the predicted and experimental values expressed in the standard unit as $\ln k_p$ and is given as:
\begin{equation}
	RMSE = \sqrt{\frac{1}{n}\sum_{i=1}^{n}\left(y_i - \hat{y_i}\right)^2}
\end{equation}
Metrics were calculated for both the training and unseen test datasets to assess the generalization of the model. Cross-validation was used during hyperparameter tuning to mitigate overfitting, and all models were trained and evaluated over multiple runs with fixed random seeds to ensure reproducibility.

\section{Results and Discussions}
\subsection{Model performance}
\begin{figure}[h]
	\centering
	\includegraphics[width=1\linewidth]{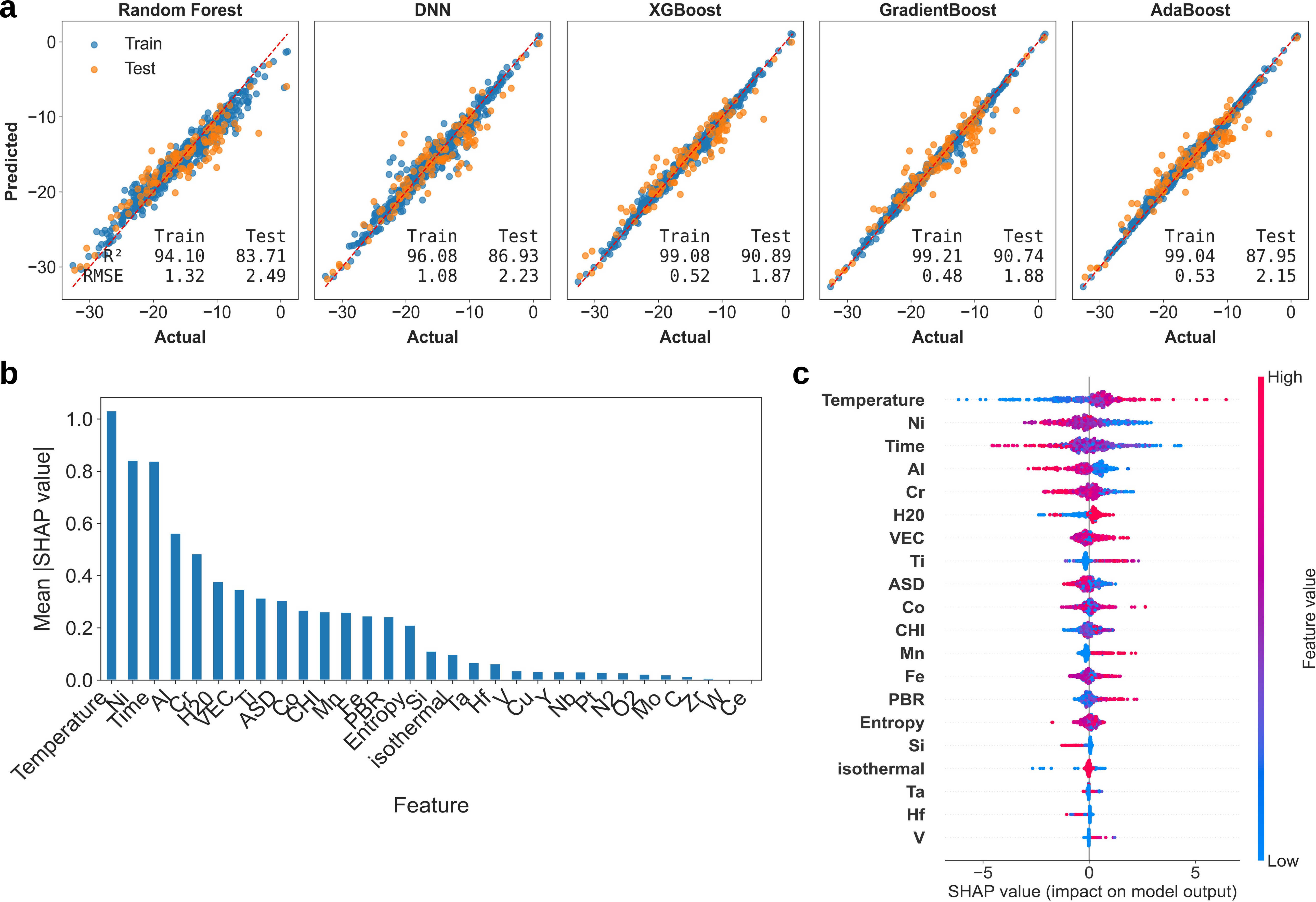}
	\caption{(a) Parity plots of the ML predicted $\ln k_p$ versus the experimental $\ln k_p$ using five different ML models; (b) SHAP summary plot of the contribution of each feature to the model; (c) SHAP dependence plot of the top 20 features that correlate with the target variable.}
	\label{fig:mod_shap}
\end{figure}

The predictive accuracy of the five machine learning models was evaluated using the test set score $\mathrm{R^2}$ and the RMSE, focusing on their ability to recover experimental trends in $\ln k_p$. From Fig. \ref{fig:mod_shap}(a), all models reproduced the experimental data with high fidelity, reflecting the strong correlation between the selected features and the oxidation kinetics. XGBoost achieved the best overall performance, with a test $\mathrm{R^2}$ of 90.89\% and RMSE of 1.87. This result explains the strength of the boosted tree ensembles in modeling highly nonlinear multivariate relationships, particularly when data come from diverse experimental sources \cite{tan2025machine}. The iterative learning process of XGBoost, which corrects residual errors of previous learners, appears especially effective in capturing interactions between descriptors \cite{tan2025machine}. GBoost performed almost identically to XGBoost ($\mathrm{R^2}$ = 90.74\%, RMSE = 1.88), reinforcing the advantage of Boosting frameworks for heterogeneous datasets. These methods appear to benefit from the structured and oxidation-descriptive feature set, which allows them to identify complex decision boundaries without overfitting, as evidenced by the close agreement between training and test performance. The DNN achieved a test $\mathrm{R^2}$ of 86. 93\% and an RMSE of 2.23. Although its accuracy was slightly lower than that of the boosting methods, the DNN still outperformed the RF and demonstrated the ability to integrate both compositional and descriptor-based inputs through its dual-branch architecture. The minor gap in performance is probably related to the dataset size relative to the model complexity, where deep learning models generally require much larger and more homogeneous datasets to fully exploit their capacity, whereas boosting methods can achieve strong generalization with fewer data points. RF and AdaBoost yielded $\mathrm{R^2}$ values of 83.71\% and 87.95\%, with RMSE values of 2.49 and 2.15, respectively. The slightly lower accuracy of RF is consistent with its fixed and non-sequential learning approach, which can limit its ability to capture fine-grained variations in the data. AdaBoost, while closer to the boosting models in accuracy, can be more sensitive to noise in heterogeneous datasets, which may explain its performance relative to XGBoost and GBoost. However, both models delivered stable predictions and minimal overfitting, confirming their suitability as robust baseline algorithms.

Table \ref{tab:model_comparison} summarizes the present results along with representative values from the literature. Li et al. \cite{li2024machine} reported that for Ni-based alloys, GBoost achieved a test $\mathrm{R^2}$ of 0.907 and an MSE of 1.422, while RF reached 0.819 and 2.767. Despite the broader compositional diversity and wider range of oxidation conditions in our dataset, our GBoost performance is comparable, and our RF result represents a clear improvement. The DNN accuracy observed here is in line with neural network results from heterogeneous datasets, although a direct comparison is made with small, single-source studies, such as in Ref. \cite{dewangan2022application}, who reported $\mathrm{R>0.999}$ for the prediction of mass gain in a single HEA system, which is not meaningful due to the controlled and homogeneous nature of these datasets.
\begin{table}[h]
    \centering
    \caption{Comparison of model performance with representative literature values for oxidation prediction of HEAs and related alloys.}
    \label{tab:model_comparison}
    \begin{threeparttable}
        \begin{tabular}{lcccc}
            \toprule
            \textbf{Model} & \multicolumn{2}{c}{\textbf{This work}} & \multicolumn{2}{c}{\textbf{Literature}} \\
            \cmidrule(lr){2-3} \cmidrule(lr){4-5}
             & $R^2$ & RMSE & $R^2$ & MSE \\
            \midrule
            RF       & 83.71 & 2.49 & 81.90\tnote{a} & 2.77\tnote{a} \\
            DNN      & 86.93 & 2.23 & 90.80\tnote{a}, 41.00\tnote{b} & 1.41\tnote{a}\;,\;1.24\tnote{b} \\
            XGBoost  & 90.89 & 1.87 & 87.20\tnote{c} & 0.17\tnote{c} \\
            GBoost   & 90.74 & 1.88 & 90.70\tnote{a}\;,\;71.00\tnote{b} & 1.42\tnote{a}\;,\;0.60\tnote{b} \\
            AdaBoost & 87.95 & 2.15 & & \\
            \bottomrule
        \end{tabular}
        \begin{tablenotes}
            \item[a] Li et al.\ \cite{li2024machine}
            \item[b] Tan et al.\ \cite{tan2025machine}
            \item[c] Gao et al.\ \cite{gao2025prediction}
        \end{tablenotes}
    \end{threeparttable}
\end{table}

The performance of gradient-boosting frameworks in our results mirrors recent findings in Ref. \cite{tan2025machine} where these methods consistently outperform deep learning when the dataset is relatively small but rich in features. In particular, XGBoost's test accuracy here exceeds the values of the reported literature for the prediction of the oxidation rate, supporting its role as a leading approach for screening oxidation-resistant HEA compositions from multisource experimental data.

\subsubsection{Feature Importance Analysis}
Feature importance analysis was performed to evaluate the contribution of each input feature to the model predictions using SHAP plots \cite{lundberg2020local2global}. As shown in Fig. \ref{fig:mod_shap}(b), temperature emerged as the most dominant factor, followed by Ni content, exposure time, and the major alloying additions Al and Cr. Several alloy design descriptors, including VEC, ASD, and $\chi$, exhibited moderate contributions, while minor alloying elements such as Hf, Ta, and V had comparatively smaller influence. The overwhelming importance of temperature reflects the Arrhenius dependence of oxidation kinetics on thermal activation, whereas the significant contributions of Ni, Al, and Cr highlight their mechanistic roles in oxide stability and growth.

\subsubsection{Interpretability Analysis of the Model}
The SHAP summary plot in Fig. \ref{fig:mod_shap}(c) provides further interpretability of the model outputs. Among the main alloying elements, both Al and Cr exerted strong protective effects: higher contents (red points) consistently shifted SHAP values to negative $\ln(k_p)$, indicating lower predicted oxidation rates. This observation is consistent with the ability of Al and Cr to form continuous $\alpha$-$\mathrm{Al_2O_3}$ and $\mathrm{Cr_2O_3}$, respectively, which are well known to suppress oxidation rates \cite{seraffon2014oxidation,chen2025effects,stott1995influence}. However, the distribution of Cr-related SHAP values also revealed a wider spread compared to Al, suggesting a somewhat context-dependent role. Although having moderate Cr content improves oxidation resistance by stabilizing the chromia scales, excessive Cr may interfere with the nucleation of alumina or promote the formation of mixed oxides, reducing the long-term protectiveness of the scale. This bidirectional influence is consistent with experimental findings that highlight Cr’s dual role in alloys, depending on composition and oxidation conditions \cite{chen2025effects,stott1995influence}. However, Ni showed the opposite trend: a higher Ni content increased $\ln(k_p)$, reflecting its tendency to form less protective NiO or destabilize Al/Cr-rich scales. The moderate influence of descriptors such as VEC, ASD, and $\chi$ supports their role in capturing the effects of electronic structure and lattice distortion on diffusion pathways and oxide adhesion \cite{tan2025machine,zeng2024machine}. Although minor alloying elements (Hf, Y) showed limited global importance in this dataset, previous studies suggest that they can provide localized improvements in scale adhesion through reactive element effects \cite{pint1996experimental,hou2023exceptional}. The SHAP analysis confirms that the model is capable of capturing both kinetic drivers such as temperature and time and composition-driven protective effects from Al and Cr.

\subsection{Screening of the NiCoCrAl compositional space}
\begin{figure}[h]
	\centering
	\includegraphics[width=0.7\linewidth]{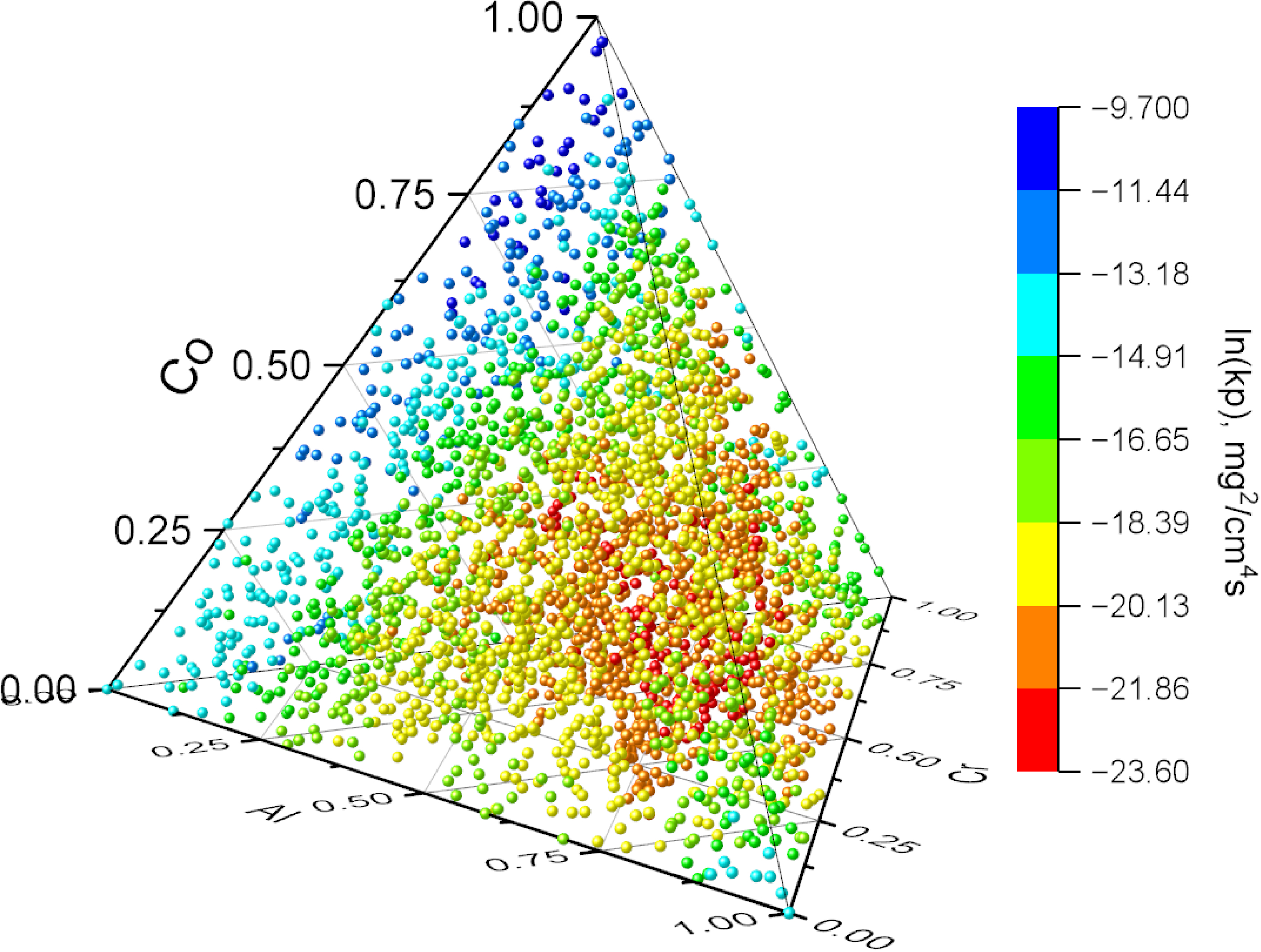}
	\caption{
		Predicted compositional dependence of oxidation resistance in the NiCoCrAl HEA system. 
		The ternary projection shows the atomic fractions of Co, Ni, Cr and Al. Data points are colored by the predicted \(\ln(k_p)\), where oxidation resistance increases with decreasing \(\ln(k_p)\) values. 
	}
	\label{fig:nicocral}
\end{figure}

The compositional map in Fig. \ref{fig:nicocral} illustrates the predicted oxidation behavior in the NiCoCrAl HEA space. Each point in the graph corresponds to a unique alloy composition defined by the atomic fractions of Co, Ni, and Cr, the Al content being implicitly determined as $C_{\mathrm{Al}} = 1 - (C_{\mathrm{Co}} + C_{\mathrm{Ni}} + C_{\mathrm{Cr}})$. The color scale represents the predicted $\ln(k_p)$, expressed in mg$^2$/cm$^4$·s. Lower values (depicted in red) indicate slower oxidation kinetics and therefore superior oxidation resistance, whereas higher values (blue) correspond to faster oxidation rates.

The distribution of colors reveals distinct compositional trends. Regions with intermediate Cr and Co contents, coupled with moderate Ni and relatively higher Al fractions, exhibit the lowest values of $\ln(k_p)$, which implies optimal oxidation resistance. This is consistent with the established role of Al in the formation of a continuous and adherent $\alpha$-Al$_2$O$_3$ scale, which provides a highly protective barrier to oxygen diffusion \cite{chen2025effects,smialek1994effects}. In contrast, compositions with an insufficient Al content tend to promote the formation of less protective Cr$_2$O$_3$ or mixed spinel oxides, which can volatilize or spall at elevated temperatures, especially in water-vapor-rich environments \cite{pint2010oxidation,seraffon2014oxidation}. The intermediate Cr levels observed in the optimal region are beneficial, as Cr assists in the initial formation of oxides and can stabilize the alumina scale through the 'third element effect', but excessive Cr can destabilize the alumina layer by preferential Cr$_2$O$_3$ growth \cite{seraffon2014oxidation,deodeshmukh2011high}.

From a mechanistic perspective, $k_p$ is related to the diffusivity of the rate-controlling species in the oxide layer through the Wagner theory of oxidation kinetics \cite{birks2006introduction}. Compositions predicted to have low values of $\ln(k_p)$ probably reduce the diffusivity of cations or anion within the oxide due to a combination of scale chemistry and microstructural factors. In NiCoCrAl alloys, higher levels of Al decrease defect concentrations in alumina, thereby lowering oxygen ion conductivity and suppressing growth rates \cite{li2019alumina}. Therefore, the predictions of the model align with thermodynamic and kinetic considerations, indicating a composition space where protective scale formation is preferred, in agreement with experimental trends reported for both superalloys and HEAs.

\subsubsection{Effect of Al and Cr on oxidation resistance}
\begin{figure}[h]
    \centering
    \includegraphics[width=1\linewidth]{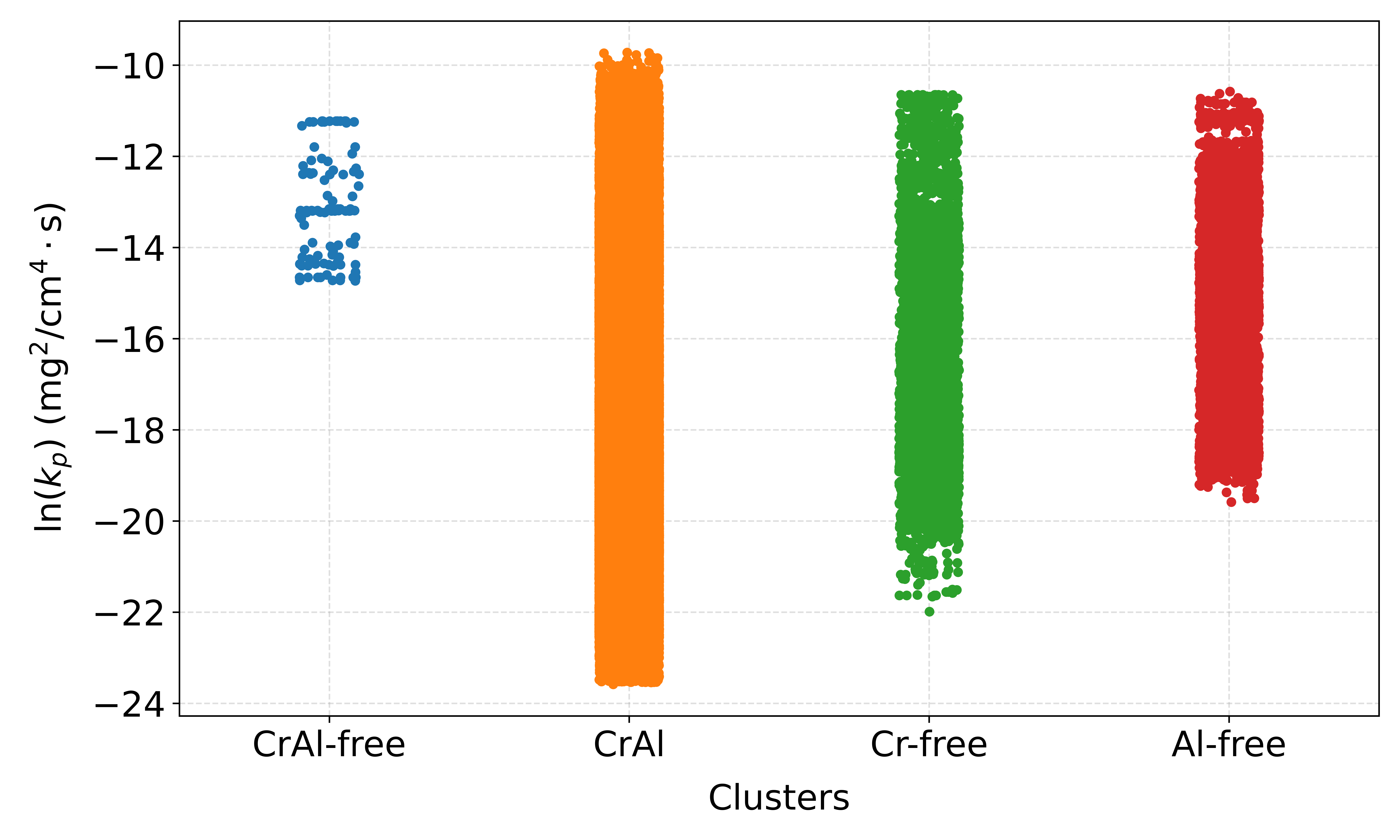}
    \caption{Predicted oxidation behavior of NiCoCrAl alloys grouped into four compositional clusters: CrAl-free, CrAl-containing (CrAl), Cr-free, and Al-free. Alloys containing both Cr and Al exhibit the lowest $\ln(k_p)$ values, highlighting their synergistic role in forming stable, protective oxide scales, whereas CrAl-free alloys show the highest oxidation rates.}
    \label{fig:al_cr}
\end{figure}

To further investigate the effect of Al and Cr on oxidation resistance, Fig. \ref{fig:al_cr} compares the predicted oxidation behavior of NiCoCrAl alloys grouped into four compositional clusters: CrAl-free, CrAl-containing (CrAl), Cr-free, and Al-free. The vertical axis shows the $\ln(k_p)$, where lower values correspond to improved resistance to oxidation. The results clearly show that alloys containing both Cr and Al (CrAl group) achieve the lowest $\ln(k_p)$ values overall, indicating the synergistic effect of Cr and Al in the formation of a protective oxide scale. In these compositions, Al promotes the development of a continuous $\alpha$-Al$_2$O$_3$ layer, while Cr supports rapid scale formation and stabilization through the “third element effect” \cite{li2022effects,niu2006third,santoro1971oxidation}. Cr-free alloys tend to perform better than Al-free alloys, indicating that the contribution of Al to oxidation resistance is more critical, consistent with the well-established role of alumina as a slow growing and adherent oxide at higher temperatures \cite{santoro1971oxidation,hou2023exceptional,deodeshmukh2011high}. In contrast, CrAl-free alloys show the highest $\ln(k_p)$ values, indicating the limited protective capacity of NiCo-based oxides without these key scale-forming elements. These trends are in agreement with thermodynamic predictions and previous experimental studies on superalloys and HEAs, where the combined presence of Cr and Al was found to substantially reduce oxidation rates at high temperatures \cite{kai2025effects,mao2025oxidation,kumar2022role}.

\subsubsection{High-temperature oxidation-resistant NiCoCrAl HEAs}
\begin{figure}
	\centering
	\includegraphics[width=1\linewidth]{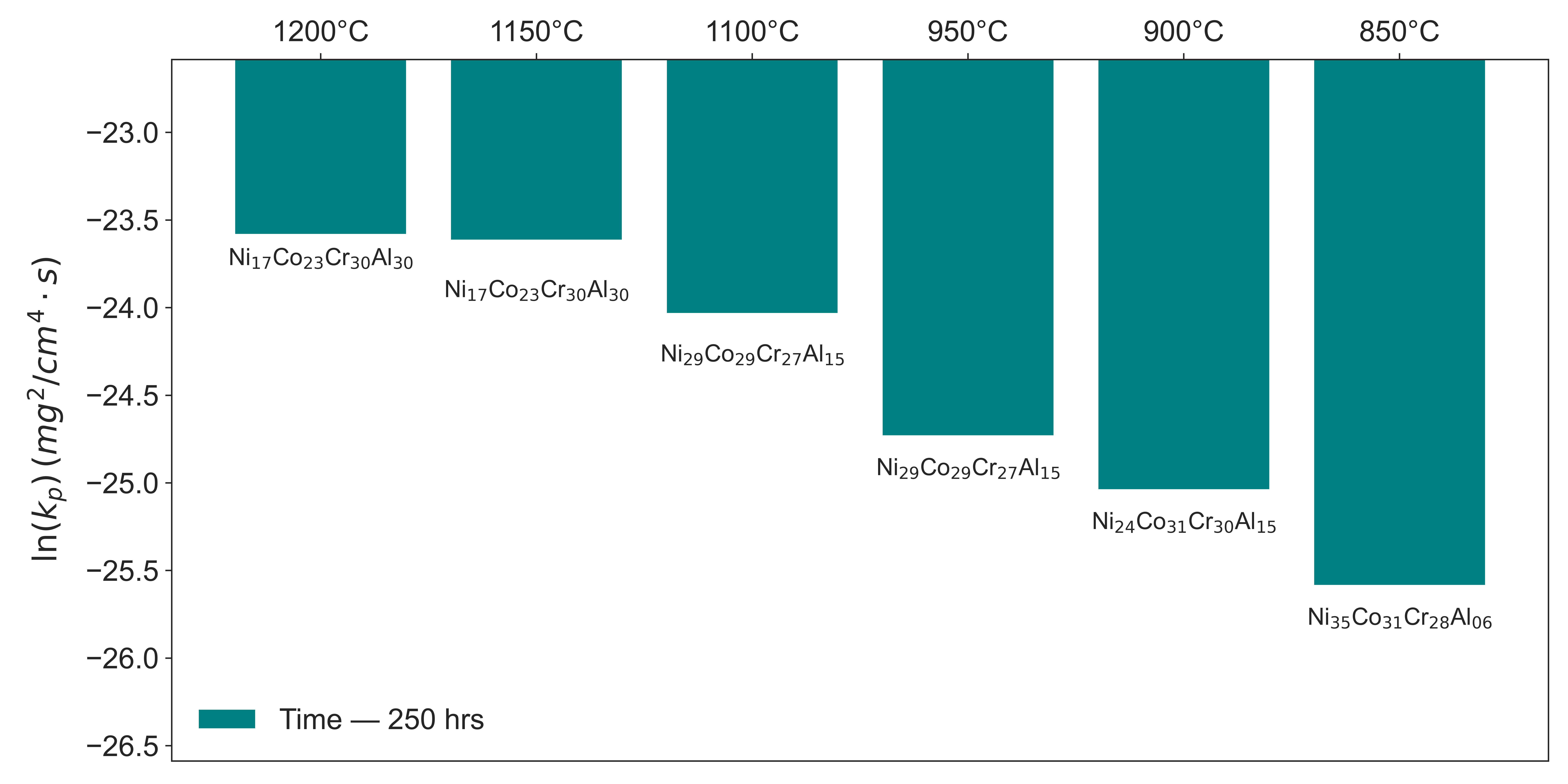}
	\caption{Predicted isothermal oxidation behaviour of NiCoCrAl alloys at various temperatures after 250 h exposure, showing the variation of $\ln(k_p)$ with composition. Lower values indicate improved oxidation resistance, with Cr–Al-rich compositions maintaining superior performance across the temperature range.}
	\label{fig:mod_pred}
\end{figure}

To find viable candidates for high-temperature oxidation-resistant bond coats, the model was used to predict NiCoCrAl HEAs over 250 h of exposure time. Fig. \ref{fig:mod_pred} shows a pronounced temperature-composition interaction. At high temperatures ($\ge$ 1150 °C), Al-rich compositions such as Ni$_{17}$Co$_{23}$Cr$_{30}$Al$_{30}$ achieve the lowest values of $\ln k_p$, consistent with the rapid establishment of a continuous $\alpha$-Al$_2$O$_3$ scale that suppresses cation and anion transport. As the temperature decreases, the optimal oxidation resistance is maintained with a reduced Al content, indicating that excessive Al is unnecessary at lower thermal loads where oxide growth kinetics are intrinsically slower. Notably, at 850 °C, a Cr-rich, low-Al alloy (Ni$_{35}$Co$_{31}$Cr$_{28}$Al$_{6}$) still exhibits very low oxidation rates, highlighting the increasing role of $\alpha$-Cr$_2$O$_3$ in providing scale protection when the kinetics of alumina formation become sluggish \cite{birks2006introduction}.

These results align with established high-temperature oxidation mechanisms for NiCoCrAl systems. The superior high-temperature performance of Al-rich alloys has been attributed to the thermodynamic stability and low growth rate of $\alpha$-Al$_2$O$_3$ \cite{heinonen2011initial,seraffon2014oxidation}, while the transition to chromia dominance at $\le$ 1000 °C has been reported in cyclic and isothermal oxidation studies \cite{barrett1975comparison,gleeson1998long}. Ni and Co form the primary solid solution matrix in NiCoCrAl alloys, providing mechanical stability and acting as a reservoir for Al and Cr to sustain the growth of protective oxide scales. In addition to this structural role, Ni and Co can participate in the formation of stable spinel oxides during high-temperature oxidation \cite{seraffon2014oxidation}. A study by Li et al. \cite{li2022effects} found that 30 at.\% Ni in a Co superalloy improved resistance to oxidation by reducing the formation of voids on the oxide scale and avoiding the spallation of the oxide layer. Spinels often develop as intermediate layers between the substrate and the alumina or chromia scale, serving as diffusion barriers and reducing oxygen diffusion \cite{zhang2024spinel,chen2015characterization}. Due to their intermediate thermal expansion coefficients, these spinels can positively reduce thermal mismatch stresses, thus improving oxide-substrate adhesion and lowering the risk of scale spallation under prolonged isothermal or cyclic conditions \cite{zhou2025superior}. Therefore, the model captures the temperature-dependent shift from alumina- to chromia-controlled protection, providing quantitative guidance for tailoring Al/Cr ratios to service conditions without excessive alloying that could compromise mechanical integrity or phase stability.

\subsubsection{Effect of reactive elements (Hf and Y)}
\begin{figure}[t]
    \centering
    \includegraphics[width=0.7\linewidth]{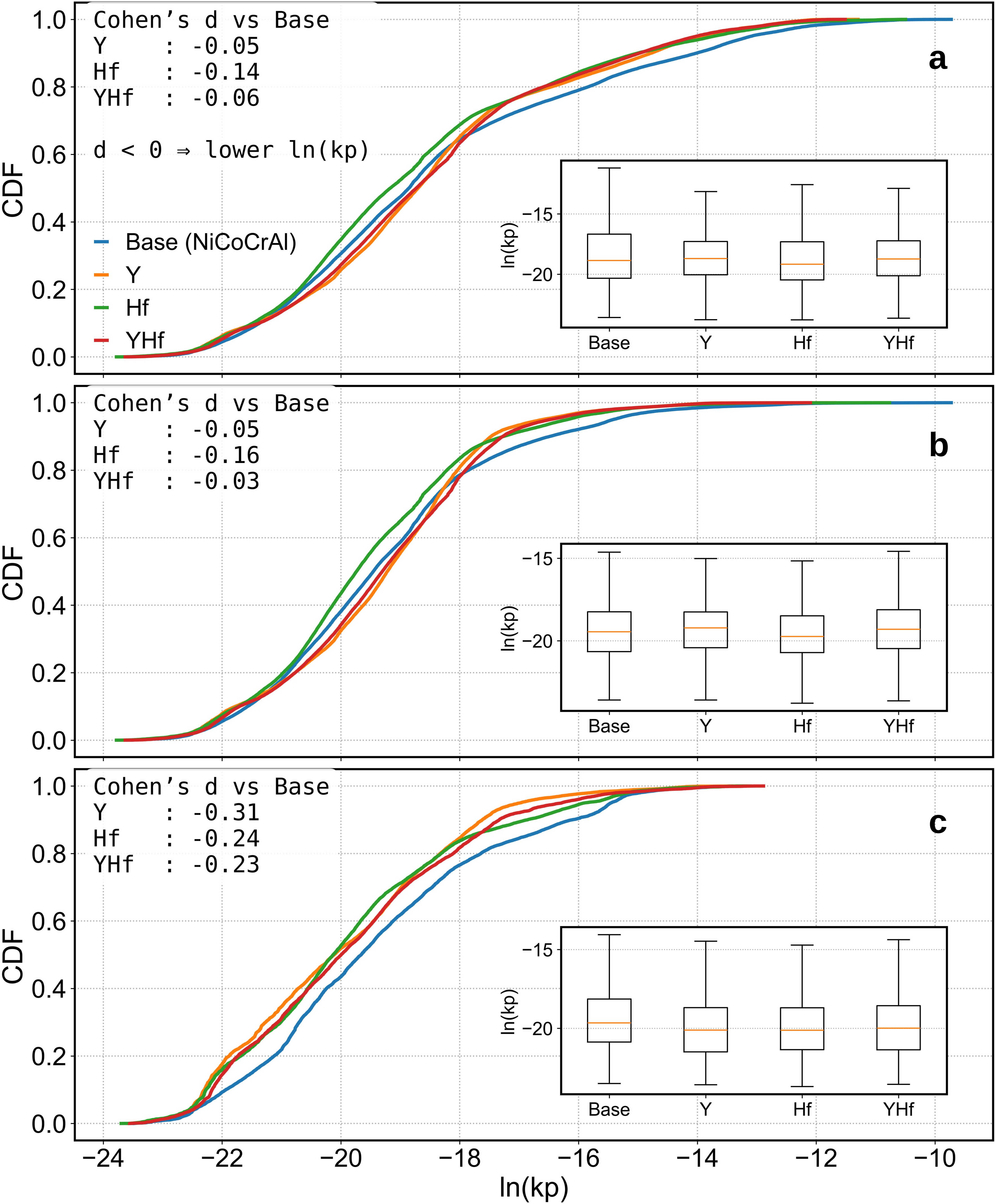}
    \caption{Effect of RE additions on the oxidation resistance of NiCoCrAl alloys. (a) NiCoCrAl base alloy; (b) NiCoCrAl with Al/Cr composition greater than 20\%; (c) NiCoCrAl alloys with NiCo composition < 30\%.}
    \label{fig:reactive}
\end{figure}

Figure \ref{fig:reactive} shows the predicted influence of Y and Hf on the oxidation resistance of NiCoCrAl alloys using cumulative distribution functions (CDF) of the ML output for $\ln k_p$. Because lower $\ln k_p$ values correspond to slower oxide growth, the curves shifted left indicate improved resistance. The base alloy NiCoCrAl exhibits the highest distribution of $\ln k_p$, while all RE modifications produce measurable leftward shifts as shown in Fig. \ref{fig:reactive}(a). The alloy containing Hf shows the highest improvement, with the median $\ln k_p$ reduced from –18.86 to –19.17 and its entire distribution displaced toward lower values. The addition of Y also reduces $\ln k_p$, but the effect is more modest (median –18.70), while the dual-doped alloy lies between at - 18.73. The inset boxplot demonstrates that REs not only lower the central tendency but also narrow the interquartile range, which leads to greater uniformity in the predicted kinetics across composition space. The effect sizes based on the distribution using Cohen's parameter $d$ confirm that Hf induces the largest shift (d = –0.14), Y induces a smaller shift (d = -0.05) and intermediate for co-doped YHf (d = –0.06). 

\begin{figure}
    \centering
    \includegraphics[width=0.7\linewidth]{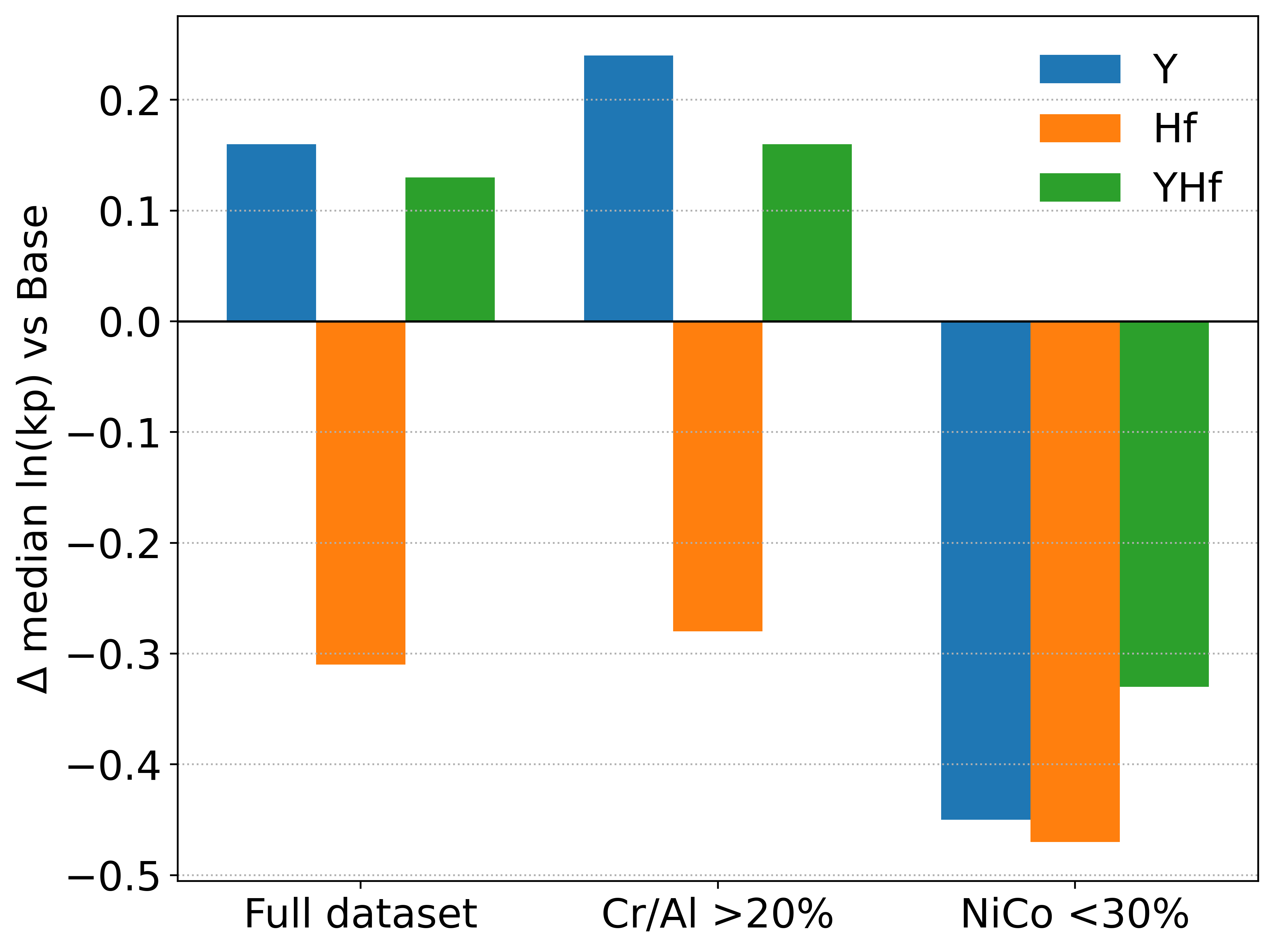}
    \caption{Median shifts of RE-added NiCoCrAl alloys with respect to base (NiCoCrAl) alloy. Notice the effect of REs on the NiCo-lean alloys.}
    \label{fig:reactive2}
\end{figure}

To understand the compositional chemistry between oxide-forming elements (Al and Cr) and RE, the Al/Cr composition of each alloy was restricted to be greater than 20\%. As shown in Fig. \ref{fig:reactive}(b), the trends become more meaningful. In Cr or Al rich alloys (Cr or Al > 20 at.\%), all CDFs move left relative to the base, reflecting the intrinsic benefit of these elements in promoting continuous chromia or alumina scales. Within this region, Hf again shows the strongest effect (median $\ln k_p$ = –19.74 compared to - 19.46 for the base), Y produces only a small change (-19.22) and YHf is between (-19.30). However, as shown in Fig. \ref{fig:reactive}(c) for NiCo-lean alloys (NiCo < 30 at.\%), the predicted behavior changes. Here, Y becomes almost as effective as Hf, with median values of -20.11 (Y) and -20.13 (Hf) essentially indistinguishable, and the effect size of Y (d = –0.31) is larger than that of Hf (d = –0.24), suggesting that Y provides the benefit of the broadest distribution in this compositional regime. Fig. \ref{fig:reactive2} condenses these results by plotting the median change in $\ln k_p$ relative to the base between regimes. The results demonstrate that Hf is the most powerful suppressor of oxidation kinetics overall, and in Cr/Al-rich alloys, Y is weaker in aggregate but becomes significant when NiCo is depleted, and YHf consistently improves resistance but does not exceed Hf alone. Consistently, the model not only reproduces the classical RE effect, where the additions of Y or Hf lower $k_p$, but also resolves its sensitivity to the composition of each alloy.

The dependence of RE effectiveness on Cr/Al enrichment and NiCo depletion is directly related to the experimental and theoretical understanding of the RE effect. Alloys with high Al or Cr contents are known to form slow-growing alumina or chromia scales and minor additions of Y or Hf segregate the oxide grain boundaries and the metal/oxide interface which reduce cation vacancy fluxes, refine the TGO grain structure, and increase interfacial toughness, mechanisms captured in dynamic segregation and oxide bonding models \cite{pint1995reactive,pint1996experimental}. As a result, TGO grows more slowly and adheres better, reducing the risk of spallation and maintaining a low effective $k_p$ \cite{birks2006introduction,pint1996experimental}. These mechanisms are widely experimentally observed in MCrAlY-type bond coats and related alloys, where the addition of Y/Hf slows scale growth and improves adhesion, often extending the life of the coating under isothermal and cyclic exposure \cite{luthra1986mechanism,smialek1994effects,carling2007effects,naumenko2016current}. The model predictions reflect this: In Cr/Al alloys, the baseline resistance is already high, and Hf magnifies the benefit by reducing $\ln k_p$ more strongly than Y. From our previous work \cite{boakye2024reactive} and experimental observations \cite{pint1995reactive}, Y is known to segregate strongly to oxide grain boundaries and interfaces at higher temperatures to prevent cation diffusion or pin sulfur in the bulk at moderate temperatures to prevent interfacial embrittlement due to sulfur segregation \cite{boakye2024reactive,holcomb2015oxidation}. The model captures the weaker effect of Y in the global dataset but becomes nearly equivalent to Hf in NiCo-lean chemistries, where its segregation-driven role is amplified. This compositional sensitivity is consistent with bond-coat studies, where Y has been shown to dramatically extend coating life under cyclic conditions by mitigating sulfur-induced spallation, particularly in Al-rich MCrAlY alloys \cite{luthra1986mechanism,carling2007effects,naumenko2016current}.

Temperature plays a crucial role in shaping these trends. At elevated temperatures, both diffusion processes and segregation kinetics are accelerated, allowing REs to rapidly reach critical interfaces and grain boundaries. Wagner’s theory of oxidation relates $k_p$ to the diffusivity of the rate-controlling species, $k_p \sim D \cdot c$, so any modification of the defect chemistry or vacancy transport can directly change $k_p$. The continuous leftward shifts observed here suggest that REs continue to reduce the effective transport coefficient even at high temperature, with no evidence of their benefit diminishing under these conditions. Beyond adhesion and segregation, additional mechanisms may also contribute. REs have been reported to alter the point defect structure of alumina and chromia, lowering the mobility of the cation vacancy, and promoting the diffusion of inward oxygen rather than outward cation flux \cite{pint1995reactive,birks2006introduction}. They also promote 'pegging' of the oxide by forming fine RE-oxide particles within the scale, which anchor the TGO to the substrate \cite{naumenko2016current,pint1996experimental}. We can argue that the absence of a synergistic kinetic effect in the dual-doped case is consistent with site-saturation arguments, such that once segregation sites at interfaces and grain boundaries are occupied and adhesion is maximized, further reductions in cation transport are limited. However, co-doping is widely recognized to improve cyclic durability by combining Y’s sulfur gettering with the adhesion benefit of Hf \cite{liu2023comparative,boakye2024reactive,huang2024superior}, even if the instantaneous $k_p$ is not lower than Hf alone. Therefore, the model correctly describes the effect of RE on compositional dependence and its persistence at high temperatures. Hf is predicted to be the most effective suppressor of oxidation kinetics, Y contributes in a subtle manner but becomes competitive in NiCo-lean and Al/Cr-rich alloys, and both act maybe through a combination of reduced transport, increased adhesion, and improved interfacial stability.

\subsection{Phase stability of model-predicted HEAs}
\begin{figure}[h]
    \centering
    \includegraphics[width=0.7\linewidth]{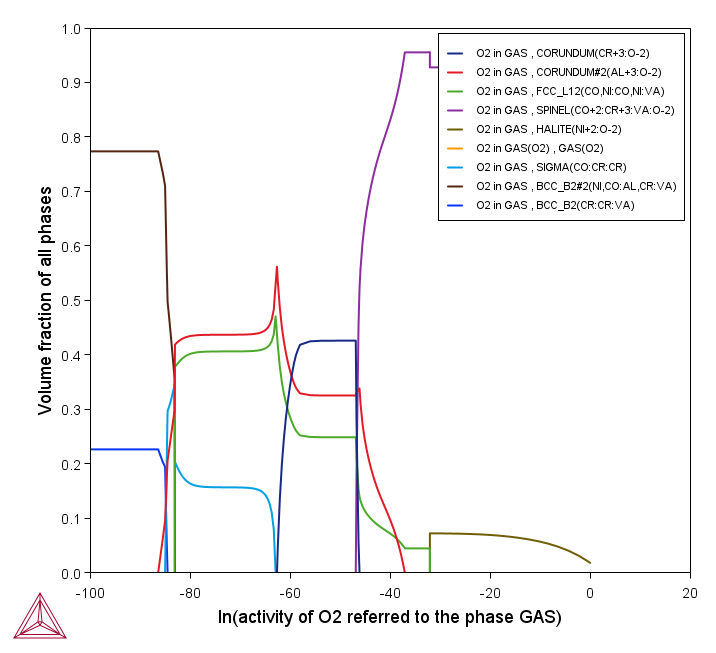}
    \caption{Phase stability of the $\mathrm{Ni_{17}Co_{23}Cr_{30}Al_{30}}$ HEA as a function of oxygen partial pressure. The plot shows the evolution of volume fractions of multiple oxide and metallic phases, including spinel, corundum, halite, and various BCC/FCC phases.}
    \label{fig:ni17o2}
\end{figure}

To further investigate the high temperature phase stability of the alloy of highest rank predicted by the model, the TCHEA8 module in Thermocalc \cite{andersson2002thermo} was used. Figure \ref{fig:ni17o2} shows the phase stability of $\mathrm{Ni_{17}Co_{23}Cr_{30}Al_{30}}$ HEA as a function of oxygen activity. The HEA undergoes a series of oxidation stages as the partial pressure of oxygen increases. Under very low pressures ($\ln a_{O_2} < -80$), the alloy remains in a mainly metallic state, stabilizing the structures of the FCC (L12), BCC (B2) and $\sigma$-phase structures. These metallic and intermetallic phases are thermodynamically stable under low oxygen activities, consistent with previous studies on oxidation-resistant HEAs \cite{tang2021high}.

As oxygen activity increases ($\approx$ -75 to -60), $\mathrm{Cr_2O_3}$ and $\mathrm{Al_2O_3}$ types and spinels begin to form. This indicates the onset of selective oxidation, where simultaneously Cr and Al are preferentially oxidized due to their higher affinity for oxygen and the negative Gibbs free energies of oxide formation, -869.3 and -555.2 kJ/mol for alumina and chromia, respectively \cite{kai2021high,hasegawa2014ellingham}. This trend has been experimentally confirmed in CoCrFeNiAlx systems, where $\mathrm{Cr_2O_3}$ and $\mathrm{Al_2O_3}$ form dense adherent scales that protect against further oxidation \cite{dkabrowa2021oxidation,veselkov2021high}. An increase in oxygen activity stabilizes the spinel phase, which dominates in the intermediate region ($\approx$ -50 to -30). The formation of spinel-type oxides is well documented in multicomponent alloy systems and has been found to contribute to both mechanical integrity and resistance to oxidation at elevated temperatures \cite{hou2023exceptional,butler2016oxidation}. At even higher pressures ($\ln a_{O_2}$ > -40), the formation of halite (NiO) becomes thermodynamically favorable, indicating late-stage oxidation of Ni. This aligns with experimental studies showing that Ni tends to oxidize only after Cr and Al are depleted or passivated \cite{chen2024selective,wang2023recent}. The predicted sequence of oxidation, from metallic phases to corundum, then spinel, and finally NiO, correlates strongly with experimental thermogravimetric and XRD observations in similar HEA systems exposed to oxidizing environments \cite{veselkov2021high}. The observed sharp phase transitions reflect thermodynamic limits rather than kinetic constraints, although real-world oxidation rates may induce smoother transitions.

\begin{figure}[h]
    \centering
    \includegraphics[width=0.5\linewidth]{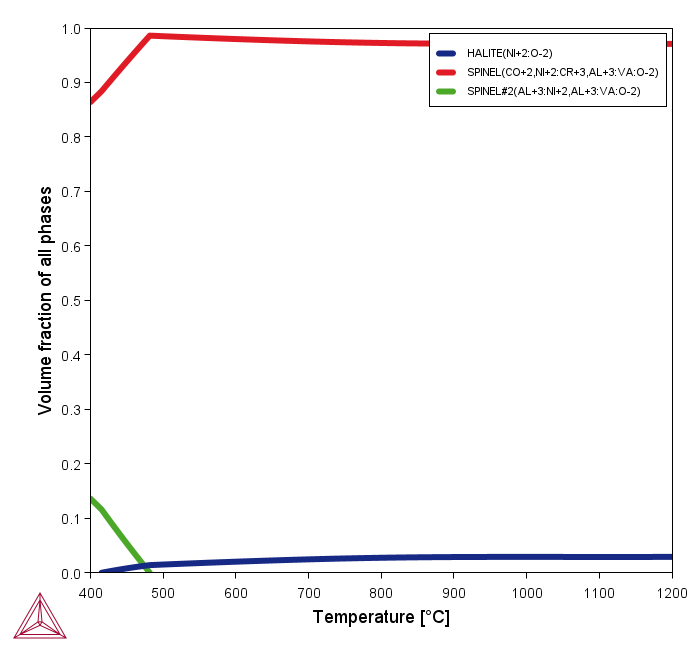}
    \caption{Volume fraction of halite and two spinel variants as a function of temperature. The spinel phase containing Co, Ni, Cr, and Al dominates, with a minor contribution from the Al-Ni-rich spinel at lower temperatures. The halite remains a minor but stable phase across the temperature range}
    \label{fig:ni17}
\end{figure}
To analyze the phase composition of $\mathrm{Ni_{17}Co_{23}Cr_{30}Al_{30}}$ in more detail, Figure \ref{fig:ni17} breaks down the evolution of the oxide phase by identifying specific variants of the spinel phase. At temperatures below 500 °C, two spinel compositions coexist mainly a Cr-Co-Ni-Al-rich spinel and an Al-Ni-rich spinel. The Al-Ni-rich spinel becomes unstable at higher temperatures and disappears, leaving Cr-Co-Al-Ni spinel as the dominant oxide. This transition suggests that the Al-Ni spinel may be metastable or stabilized kinetically at low temperatures but is thermodynamically less favorable at elevated temperatures. A similar phenomenon was reported in Al-containing CoCrFeNi alloys, where low temperature oxidation resulted in the formation of a complex spinel and mixed oxide, which simplified into a single-phase spinel at higher temperatures \cite{korda2023high}. The halite phase (NiO) remains a minor phase but gradually increases with temperature, which is consistent with the late stage oxidation of Ni observed in experimental work on Ni-containing alloys \cite{birks2006introduction,korda2023high}.

\subsection{Model limitations and future work}
Although the present study achieves good accuracy and provides information on the oxidation behavior of NiCoCrAl HEAs, several limitations must be acknowledged. First, the model relies on 743 experimental measurements collected from various sources in the literature. Although the dataset consists of a wide range of elemental compositions and oxidation conditions, it remains inadequate compared to that of the vast compositional space of HEAs. In perspective, specific HEA chemistries and service environments, such as those performed under cyclic oxidation, are still underrepresented, which limits the model's ability to confidently extrapolate uncharted regions. Second, $k_p$ was adopted as the only metric of oxidation resistance. Although the measure of oxidation kinetics is widely explained by $k_p$, it does not account for other degradation mechanisms. These include stress-induced spallation, sulfur segregation at grain boundaries or interfaces, and microcracking during thermal cycling, which often dictate the long-term performance of bond coats. Third, the model does not explicitly incorporate microstructural factors, including grain boundaries, defect densities, or phase morphology; however, these features strongly influence oxide growth and scale adhesion. The predictions represent nominally homogeneous alloys and may not fully account for processing-induced variability. Finally, while the inclusion of REs is a key step, the beneficial concentrations of these elements are not yet well constrained because of limited experimental data. Overdoping effects, which can accelerate oxidation or embrittle the coating, remain beyond the resolution of the current framework.

These limitations also point to opportunities for future development. The expansion of the dataset with systematic cyclic oxidation experiments and underrepresented compositions will enhance generalizability. Integrating microstructural descriptors from advanced characterization techniques or first-principles simulations would allow the model to capture processing-structure-property relationships more faithfully. Coupling the ML framework with diffusion kinetics and mechanistic models of transient oxide formation could also provide a more holistic description of service degradation. Furthermore, expanding the scope of thermodynamic validation beyond equilibrium phase stability to include kinetic simulations would strengthen confidence in the predicted alloy chemistries. 

\section{Conclusion}
This study demonstrates the potential of a machine learning model as a powerful data-driven tool for accelerating the discovery of oxidation-resistant HEA coatings. By integrating a high-fidelity experimental dataset with physical principles, the model predicts with high precision the parabolic oxidation constant over a wide compositional range, outperforming or matching the state-of-the-art models in the literature. The interpretation of the model reveals trends that are consistent with the critical role of the Al-Cr synergy in the formation of stable, protective oxide scales, the temperature-dependent shift between alumina and chromia control, and the measurable benefits of RE additions such as Y and Hf in improving scale adhesion and slowing growth kinetics.

High-throughput compositional screening identifies NiCoCrAl-based HEAs that combine optimal oxidation resistance with favorable phase stability through Thermo-Calc analysis. These findings offer clear and quantitative guidance for tailoring compositions to specific service conditions, while minimizing the reliance on costly and time-consuming trial-and-error experimentation. The model is not limited to NiCoCrAl HEAs but also provides a generalizable framework enabling rapid alloy design in a wide range of high-temperature applications.

\section*{Acknowledgments}
This research was supported by the NSERC Alliance International Catalyst (ALLRP 592696-24), Canada, and the use of a high-performance computing system at the University of Manitoba and the Research Alliances of Canada. The authors thank Ana Hernandez, of Thermo-Calc Software, for assistance in generating the phase transformation diagrams using Thermo-Calc. 

\section*{CRediT authorship contribution statement}
\textbf{Dennis Boakye}: Writing – review and editing, Writing – original draft, Visualization, Validation, Methodology, Investigation, Formal analysis, Data curation. \textbf{Chuang Deng}: Writing – review and editing, supervision, software, resources, project administration, investigation, Fund Acquisition, conceptualization.

\section*{Declarations}
The authors declare that they have no known competing financial interests or personal relationships that could have appeared to influence the work reported in this paper.

\section*{Data availability}
The raw dataset used for the study is attached as Supplemental file 2. The study code is available on reasonable request.

\section*{Supplementary information}
The supplementary information referenced in the main text is attached as the Supplemental file.



\small
\bibliographystyle{elsarticle-num} 
\biboptions{sort&compress}
\bibliography{references}





\end{document}